# Suppression of Edge Recombination in InAs/InGaAs DWELL Solar Cells

Tingyi Gu, Mohamed A. El-Emawy, Kai Yang, Andreas Stintz, and Luke F. Lester
Center for High Technology Materials, University of New Mexico, 1313 Goddard SE, Albuquerque, NM 87106,U.S.A

## ABSTRACT

The InAs/InGaAs DWELL solar cell grown by MBE is a standard *pin* diode structure with six layers of InAs QDs embedded in InGaAs quantum wells placed within a 200-nm intrinsic GaAs region. The GaAs control wafer consists of the same *pin* configuration but without the DWELL structure. The typical DWELL solar cell exhibits higher short current density while maintaining nearly the same open-circuit voltage for different scales, and the advantage of higher short current density is more obvious in the smaller cells. In contrast, the smaller size cells, which have a higher perimeter to area ratio, make edge recombination current dominant in the GaAs control cells, and thus their open circuit voltage and efficiency severely degrade. The open-circuit voltage and efficiency under AM1.5G of the GaAs control cell decrease from 0.914V and 8.85% to 0.834V and 7.41%, respectively, as the size shrinks from $5*5mm^2$ to $2*2mm^2$, compared to the increase from 0.665V and 7.04% to 0.675V and 8.17%, respectively, in the DWELL solar cells.

The lower open-circuit voltage in the smaller GaAs control cells is caused by strong Shockley-Read-Hall (SRH) recombination on the perimeter, which leads to a shoulder in the semi-logarithmic dark IV curve. However, despite the fact that the DWELL and GaAs control cells were processed simultaneously, the shoulders on the dark IV curve disappear in all the DWELL cells over the whole processed wafer. As has been discussed in previous research on transport in QDs, it is believed that the DWELL cells inhibit lateral diffusion current and thus edge recombination by collection first in the InGaAs quantum well and then trapping in the embedded InAs dots. This conclusion is further supported by the almost constant current densities of the different area DWELL devices as a function of voltage.

## INTRODUCTION

Recent interest in using InAs quantum dots (QDs) in the absorbing region of solar cells has focused primarily on the predicted increase in quantum efficiency due to the intermediate band effect or simply larger short circuit current density [1-4]. However, the three-dimensional carrier confinement inherent to QDs endows them with unique carrier transport capabilities that have not been previously explored in the context of solar cells. In this work, it is observed that InAs/InGaAs "dots-in-a-well" (DWELL) structures [5-7] efficiently suppress lateral carrier diffusion. Therefore, not only do the DWELL structures enhance photocurrent by extending the absorption edge, but they should also inhibit the spreading of current to the perimeter of a device where edge recombination can dominate [8-10]. In this paper, we examine this premise by comparing the dark current behavior of DWELL cells and GaAs control cells of varying area. The results are promising for applications such as concentration and flexible surfaces where shrinking the size of the device while maintaining high charge collection efficiency are of paramount importance.

**EXPERIMENT**

The control and DWELL samples were fabricated simultaneously to minimize process variation. The Ge/Au/Ni/Au emitter metallization creates the solar cell finger grid and is laid out in three different areal dimensions (5×5mm$^2$, 3×3mm$^2$, and 2×2mm$^2$). The bottom Ti/Pt/Au p-type contact is common for the solar cells on the sample. A 270-nm deep mesa, which reaches the intrinsic region, is dry-etched to separate neighboring solar cells with an isolation resistance of ~10$^5$ Ω. Finally, an anti-reflective coating (ARC) layer is deposited on the front surface for reducing the reflection loss and improving the surface passivation. The ARC layer is 80-nm thick Si$_x$N$_y$ with a refractive index around the geometric mean of air and GaAs.

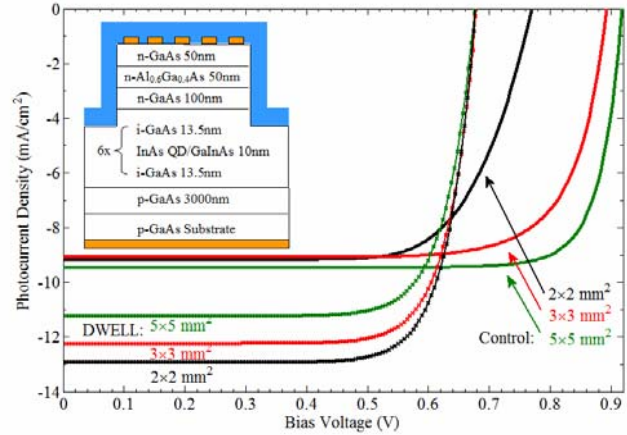

**Figure 1:** Photocurrent of DWELL and GaAs control cell of different sizes (2×2 mm$^2$, 3×3mm$^2$ and 5×5 mm$^2$) under AM 1.5G illumination. The inserted picture is the schematic diagram of the DWELL solar cell with six-stacks of InAs QDs embedded in InGaAs quantum wells.

For IV characterization, the cell is illuminated using an ABET Technologies 150-Watt Xe lamp. A filter is inserted between source and cell to simulate the AM1.5G spectrum. The solar cell is connected to an HP parameter analyzer by a four-point probe approach to eliminate the series resistance introduced by the probes and the parameter analyzer. A TE cooler is used to fix the cell temperature at 25.0±0.1°C throughout the test. As shown in Fig. 1, the typical DWELL device exhibits higher short circuit current density ($J_{SC}$) while maintaining the same open circuit voltage ($V_{OC}$) for smaller areas. For the GaAs control cells, however, smaller size, which has a higher perimeter-to-area ratio, makes edge recombination current dominant in these devices, and, thus, severely impacts their $V_{OC}$ and efficiency. Here $V_{OC}$ of the 2×2 mm$^2$ GaAs cell is 10% lower than the 5×5 mm$^2$ one as shown in Fig. 1 and table 1.

**Table 1**: Measured short circuit current densities ($J_{sc}$), open circuit voltages ($V_{oc}$), and efficiencies of the GaAs control cells and InAs DWELL solar cells under AM1.5G illumination.

| Size | $J_{sc}$ (mA/cm$^2$) | | $V_{oc}$ (V) | | Efficiency (%) | |
|---|---|---|---|---|---|---|
| | Control | DWELL | Control | DWELL | Control | DWELL |
| 5×5 mm$^2$ | 9.46 | 11.23 | 0.914 | 0.665 | 8.85 | 7.04 |
| 3×3 mm$^2$ | 9.08 | 12.23 | 0.890 | 0.670 | 7.61 | 7.79 |
| 2×2 mm$^2$ | 9.17 | 12.93 | 0.834 | 0.675 | 7.41 | 8.17 |

## MODELING

To investigate the underlying physics of the $V_{OC}$ degradation in the control samples, the dark IV is measured and the carrier recombination mechanism is analyzed. Here, the conventional single-diode model described in [11] with constant reverse saturation current and ideality factor fails to describe the dark behavior of either the control or DWELL cells, so different models involving non-radiative recombination on the edge or in the quantum dot layers are constructed for the control and DWELL cells, respectively.

The ideality factors for both the control and DWELL cells are measured as shown in Fig. 2(a) and (c). Substantial differences between the GaAs and DWELL cells include the shoulder in the GaAs cells' IV curves and the resulting hump in the local ideality factor. Neither of these effects is observed for any area size in the DWELL cells. The peak in the ideality factor is more significant as the area of the GaAs cell decreases, which suggests that edge recombination is important. Another series of wafer growths and processing produced the same results. This strongly voltage-dependent ideality factor can be modeled by the pinning of the Fermi-level to surface states at the device perimeter [12-15].

### **GaAs control cell modeling**

The relevant equations for modeling the GaAs controls cells are:

$$J_d = J_b + J_p \tag{1-a}$$

$$J_b = J_{b0} \exp\left(\frac{V - J_d \times A \times R_s}{n_b V_t}\right) \tag{1-b}$$

$$J_p = q \frac{n_s p_s - n_i^2}{(n_s + n_1)/S_{p0} + (p_s + p_1)/S_{n0}} \times d \times \frac{P}{A} \tag{1-c}$$

where $J_b$ is the bulk contribution and $J_p$ is from the perimeter of the cell. The parameters are well adjusted for this model to fit the experimental data as shown Fig. 2(a) and (b). $J_b$ follows the conventional diode equation, and $J_p$ is modeled using Shockley-Read-Hall (SRH) statistics as expressed in equation (1-c) [16]. It is assumed that the ideality factor ($n_b$) and reverse saturation current density ($J_{b0}$) of the bulk diode are constants over the bias range where SRH recombination dominates. The surface carrier density ($p_{s0}$, $n_{s0}$) influences the peak location of the hump in the ideality factor in Fig. 2(a), and the surface recombination rate ($S_{p0}$, $S_{n0}$) determines the shape of the hump. At high bias (>0.8V), the series resistance ($R_s$) dominates the trend. Based on these features, the model is adjusted to fit the tested ideality factor and dark current density (Fig. 2 (a-b)). The parameters used in equation (1-b) for describing the bulk component, are the same for three different scales, while the exposed edge surface to area ratio is rising with the shrinking size. The misfit between the model (dash line) and the experiment (solid line) in Fig. 2 (a) and (b) might be due to the non-uniform current distribution as shown in the Silvaco simulation picture inserted in Fig. 2(c).

In equation (1), $n_i$ is the intrinsic carrier density for GaAs, and $V_t$ is the thermal voltage at room temperature. $p_s, n_s = p_{s0}, n_{s0} + dn$ where $dn$ is the injected carrier density ($n_i exp(V/2/V_t)$). $n_b$ is 1.31, $J_{b0}$ is $1.2*10^{-10}$ mA/cm$^2$. $S_{p0}$ is 0.8, 1.0, 1.0x10$^7$cm/s, $S_{n0}$=7, 3, 2x10$^7$cm/s, $p_{s0}$ is 6, 1, 3x10$^{13}$ cm/s, $R_s$ is 2.2, 1.0, 0.6 Ohm, and the exposed edge surface to diode area ratio is 10, 5.5, 2.5x10$^{-6}$ for the 2x2, 3x3 and 5x5 mm$^2$ cells, respectively.

### **DWELL cell modeling**

Here, the dual diode model is applied to simulate the dark behavior of the DWELL cells:

$$J_d = J_{diff} + J_{rec} \quad (2\text{-}a)$$

$$J_{diff} = J_{01}\left[\exp\left(\frac{V - J_d \times A \times R_s}{n_1 V_t}\right) - 1\right] \quad (2\text{-}b)$$

$$J_{rec} = J_{02}\left[\exp\left(\frac{V - J_d \times A \times R_s}{n_2 V_t}\right) - 1\right] \quad (2\text{-}c)$$

where the dark current is decomposed into the diffusion ($J_{diff}$) and recombination ($J_{rec}$) parts. The diffusion part from the bulk is the same as in the GaAs control (equation 1-b), but the edge component is adjusted from SRH statistics to treat the nonradiative recombination current in quantum dots with constant ideality factor ($n_2$) and reverse saturation current ($J_{02}$). The parameters are the same for the three different scales, where $J_{01}$ equals $J_{b0}$, $n_1$ equals $n_b$, $J_{02} = 7 \times 10^{-8}$ mA/cm$^2$, and $n_2$ is 2.

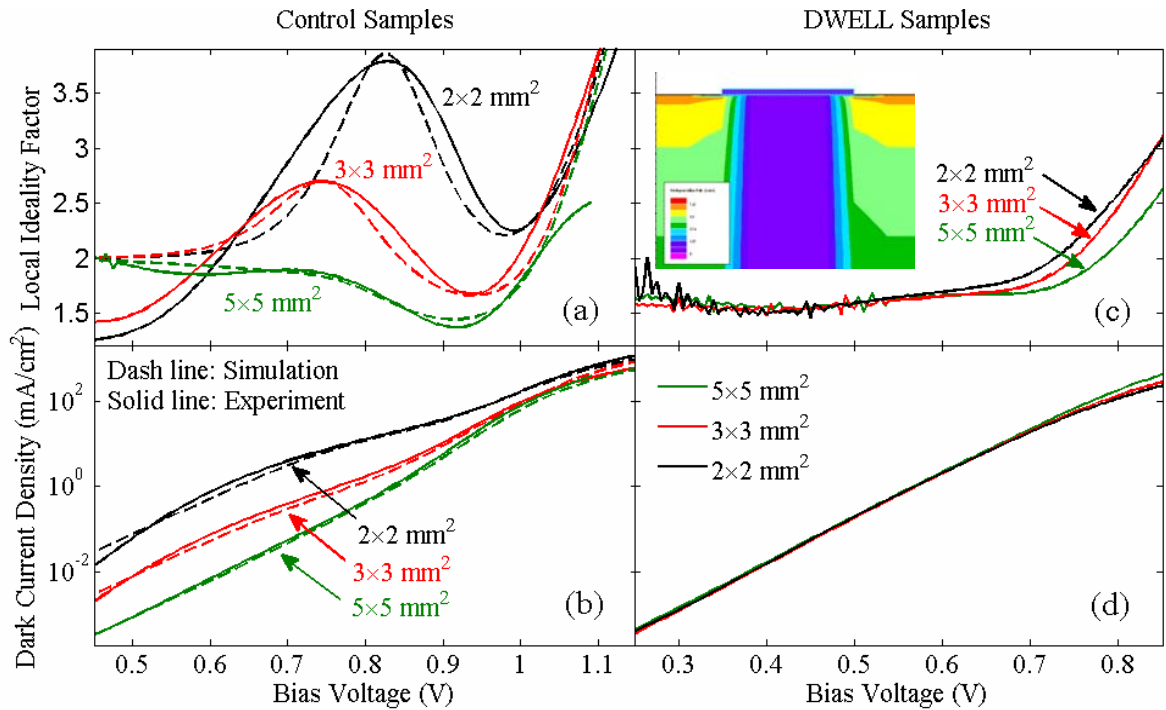

**Figure 2:** A comparison of the dark behavior of GaAs control and DWELL cells for the same dimensions (2×2 mm$^2$, 3×3 mm$^2$ and 5×5 mm$^2$). Measured and simulated local ideality factor (a), and the measured and simulated semi-logarithmic dark current density (b) for the control cells. Measured local ideality factor (c) and dark current density (d) for the DWELL cells. The simulation is based on Equation 1(a), where the parameters extracted by the curve fitting are illustrated in the modeling part. The inserted picture is a Silvaco simulation of the non-uniform current distribution in the device.

## DISCUSSION

Good agreement is achieved between the model and the data for Figs. 2(a) and (b). It is found that the edge recombination current is proportional to the perimeter of the cell, while the bulk current scales with the cell area. Therefore, as predicted by the simulation and confirmed experimentally, the smaller cells, which have a comparatively larger P/A ratio, are more susceptible to the edge recombination phenomenon. Any minor disagreement between the experiment and model can be explained by our assumption that there is uniform edge recombination current across the device perimeter and that $n_b$ and $J_{b0}$ are constants. The edge recombination component has been simplified to a 1D model with constant etched depth and surface states over the exposed perimeter. In reality, however, the recombination current is most intense near the contact fingers and decreases with distance away from the metal edges. This was verified by 2D electroluminescence of the device and a SILVACO ATLAS simulation.

Although the DWELL and GaAs control cell were processed in the same run, the humps in the ideality factor disappear completely in all of the DWELL cells as demonstrated in Fig. 2(c). Similar to previously published observations [8], the DWELL structure is effective at blocking lateral current flow to the device perimeter where surface recombination can occur. Although thermal re-emission and non-radiative recombination generally increase the dark current of the DWELL cells compared to the control ones, the overlapping IV curves shown in Fig. 2(d) for different size DWELL devices further supports the idea that the dots play an effective role in suppressing edge current.

## CONCLUSIONS

In summary, compared to GaAs *pin* diode cells that experimentally display degradation of the dark current and ideality factor as the device perimeter/area ratio is increased, solar cells with an InAs/InGaAs DWELL structure positioned in the intrinsic region do not exhibit this problem. The strong peaking of the ideality factor in the GaAs control cells has been theoretically explained by a model that includes bulk and edge recombination effects. Since a hump in the ideality factor of the DWELL cell is completely absent, it is concluded that the DWELL structure limits lateral current movement and subsequent edge recombination. The DWELL devices should be especially useful for concentrated and flexible solar applications for which small area devices are highly desirable.

## ACKNOWLEDGMENTS

This work was supported in part by Dr. Kitt Reinhardt of the Air Force Office of Scientific Research under grant number FA9550-06-1-0407 and the Micro-Autonomous Systems Technology (MAST) program administered by the US Army Research Lab.